\documentclass[prb,showpacs,floatfix,twocolumn]{revtex4}

\usepackage{graphicx} 

\def\cio{Cr:In$_2$O$_3$}
\def\io{In$_2$O$_3$}

\begin{document}

\title{Ferromagnetism and spin polarized charge carriers in In$_{2}$O$_{3}$\, thin films}
\author{Raghava P. Panguluri$^{a}$, P. Kharel$^{a}$, C. Sudakar$^{a}$, R.
Naik$^{a}$, R. Suryanarayanan$^{a\dagger }$, V.M. Naik$^{b}$, A.G. Petukhov $%
^{c}$, B. Nadgorny$^{a}$, G. Lawes$^{a}$}
\affiliation{$^a$Department of Physics and Astronomy, Wayne State University, Detroit, MI
48201\\
$^{b}$Department of Natural Sciences, University of Michigan-Dearborn,
Dearborn, MI 48128}
\affiliation{\\
$^{c}$Department of Physics, South Dakota School of Mines, Rapid City, SD
57701\\
$\dagger $ Permanent address: LPCES, CNRS, ICMMO, Universite Paris-Sud,
91405 Orsay, France}
\date{\today }

\begin{abstract}
We present evidence for spin polarized charge carriers in \io\, films.  Both
\io\, and Cr doped \io\, films exhibit room temperature ferromagnetism after
vacuum annealing, with a saturation moment of approximately 0.5 emu/cm$^3$.  
We used Point Contact Andreev Reflection measurements to directly determine
the  spin polarization, which was found to be approximately 50$\pm$5\% for
both compositions.  These results are consistent with suggestions that the
ferromagnetism observed in certain oxide semiconductors may be carrier mediated.
\end{abstract}


\maketitle

The potential technological applications of magnetic semiconductors to the
field of spintronics have motivated the study of many promising systems,
including (Ga,Mn)As \cite{macdonald,R1} and transition metal doped
semiconducting oxides (DMSO)\cite{coeyB,chambers}. Some of these latter
systems have been predicted to exhibit room temperature ferromagnetism\cite%
{ohno}, which has been observed experimentally in Co doped TiO$_{2}$\cite%
{cotio}, Co doped ZnO\cite{cozno}, and Cr doped \io \cite{moodera}.
However, the origin of ferromagnetism in these DMSO materials remains enigmatic,
in part because of the possibility of a magnetic signal arising from
undetected transition metal oxide impurity phases\cite{impurity}. At the same time,
ferromagnetism has been observed in a number of undoped oxide
samples including HfO$_{2}$\cite{hfo2}, TiO$_{2}$\cite{tio2}, 
In$_{2}$O$_{3}$\cite{tio2}.  The recent
results on ferromagnetism in carbon\cite{carbon} emphasizes the importance of vacancies and other defects
in promoting ferromagnetic order.  

In$_{2}$O$_{3}$ is a transparent
semiconductor and can be highly conductive at
room temperature when doped, making ferromagnetic In$_{2}$O$_{3}$ films
attractive candidates for magneto-optical and spintronic devices. Room
temperature ferromagnetism has been predicted for Mo doped In$_{2}$O$_{3}$
films \cite{moio} and observed in Ni, Fe, and Co doped samples \cite{tmdoped}
as well as undoped In$_{2}$O$_{3}$\cite{tio2}. It has been shown that the
electrical and magnetic properties of Cr:In$_{2}$O$_{3}$\thinspace\ films
are both sensitive to the oxygen vacancy defect concentration, and that the
ferromagnetic interaction depends on carrier density\cite{moodera}. While it
has been suggested that ferromagnetism in \cio\, films is carrier mediated\cite{zunger3},
the precise relationship between the spin transport properties of the charge
carriers and the net ferromagnetic moment remains unclear. 

In this Letter
we demonstrate that the charge carriers in
undoped In$_{2}$O$_{3}$ films have a significant spin
polarization at helium temperatures. Furthermore, measurements 
on Cr doped \io\, samples yield quantitatively similar results to measurements on undoped
samples, suggesting that transition metal dopants may not play any significant
role in the development of ferromagnetic order. 

We prepared ceramic samples of In$_{2}$O$_{3}$\thinspace\ (with a base
purity of 99.99\%) and In$_{2}$O$_{3}$\thinspace\ doped with 2 at\% Cr using
a standard solid state process\cite{kharelcio}. The powder samples were
pressed into 2'' diameter sputtering targets, then annealed in air at 1100 $%
^{\circ }$C for 6 hours. In$_{2}$O$_{3}$\thinspace\ and Cr:In$_{2}$O$_{3}$%
\thinspace\ thin films were deposited by reactive magnetron sputtering of
this target using an RF power source. High-purity argon was used as the
sputtering gas and a small partial pressure of oxygen was maintained to
obtain stoichiometric films. Oxygen at a partial pressure of 10$^{-3}$ torr
and argon at a partial pressure of 1.4x10$^{-2}$ torr were used as reactive
and sputtering gases respectively. The films were deposited onto (0001)
oriented single crystal sapphire substrates. While the as-prepared
samples were insulating and non-magnetic, the films became conducting and
ferromagnetic when annealed in vacuum for 6-8 hours.

\begin{figure} \smallskip \centering 
\includegraphics[width=70mm]{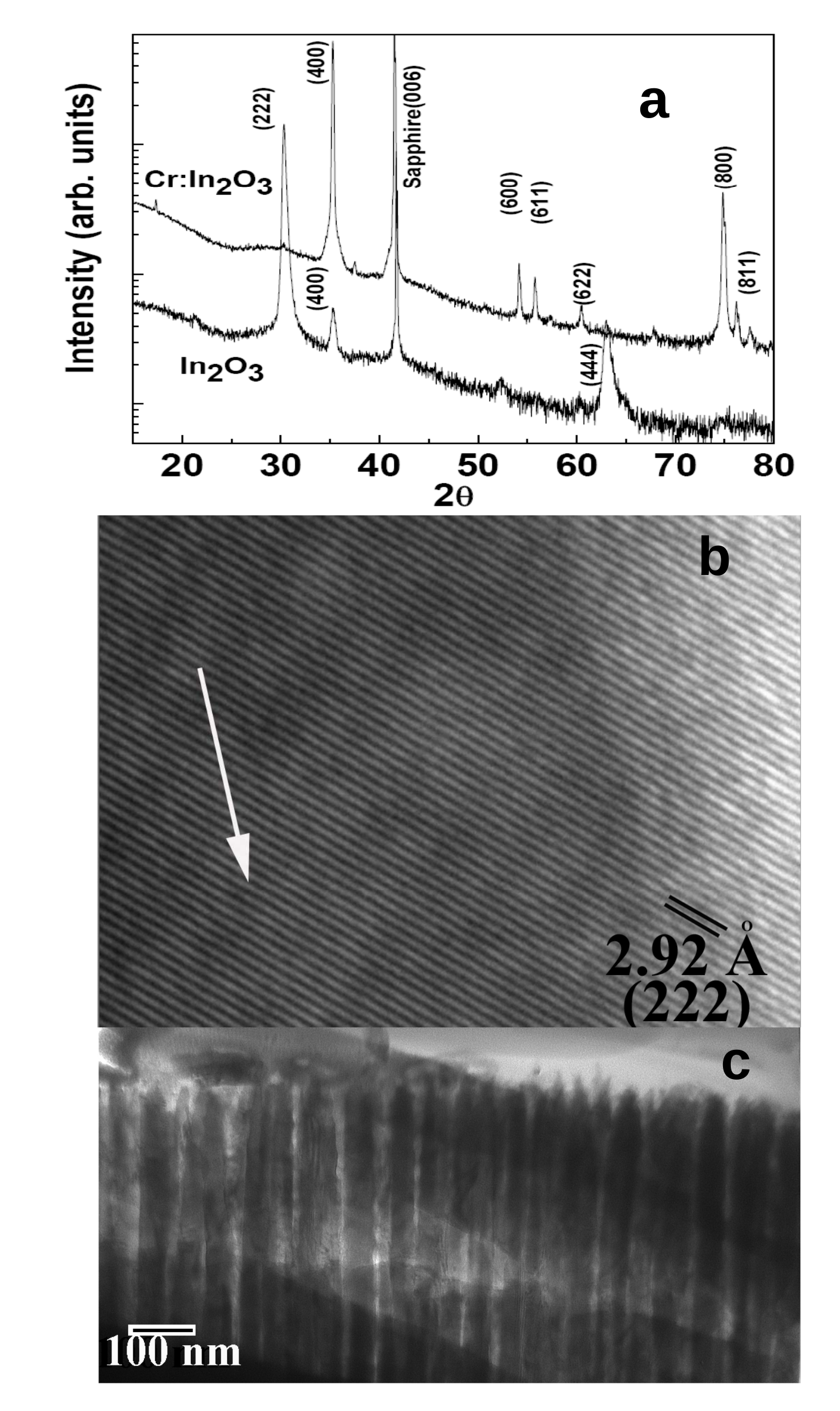}
\caption{(a) X-ray diffraction
spectrum for a pure \io\, sample (upper curve) and a Cr doped to 2 at.\% In$_2$O$_3$\, thin film (lower curve).
 (b) HRTEM of the Cr:In$_2$O$_3$\, thin film, showing the
crystalline structure and absence of any impurity phase. This arrow
indicates the direction of the columnar axis. (c) Cross-sectional TEM image
showing the highly textured thin film.}
\label{fig:XRD}
\end{figure}

 The X-Ray Diffraction
(XRD) spectra for the \io\, and Cr:In$_{2}$O$_{3}$\thinspace\ samples,
are shown in Figure \ref{fig:XRD}a. The polycrystalline films are
textured, with strong diffraction peaks indicating a preferred orientation along (222) or (400). There is
no evidence for secondary phase formation.  We show high resolution (HR) and
cross-sectional transmission electron microscope (TEM) images of a Cr:In$%
_{2} $O$_{3}$\thinspace\ film in Figs \ref{fig:XRD}b and \ref{fig:XRD}c. The
HRTEM image shows the absence of defects, secondary phases, or clusters in
these high-quality samples. Extensive SEM EDS
mapping of the In$_{2}$O$_{3}$\thinspace\ films (not shown) gave no indication of any
transition metal dopants, including Cr, Co, Fe, Ni, and Mn, thus ruling out the
possibility of accidental contamination with magnetic transition metal
impurities.

\begin{figure} \smallskip \centering 
\includegraphics[width=70mm]{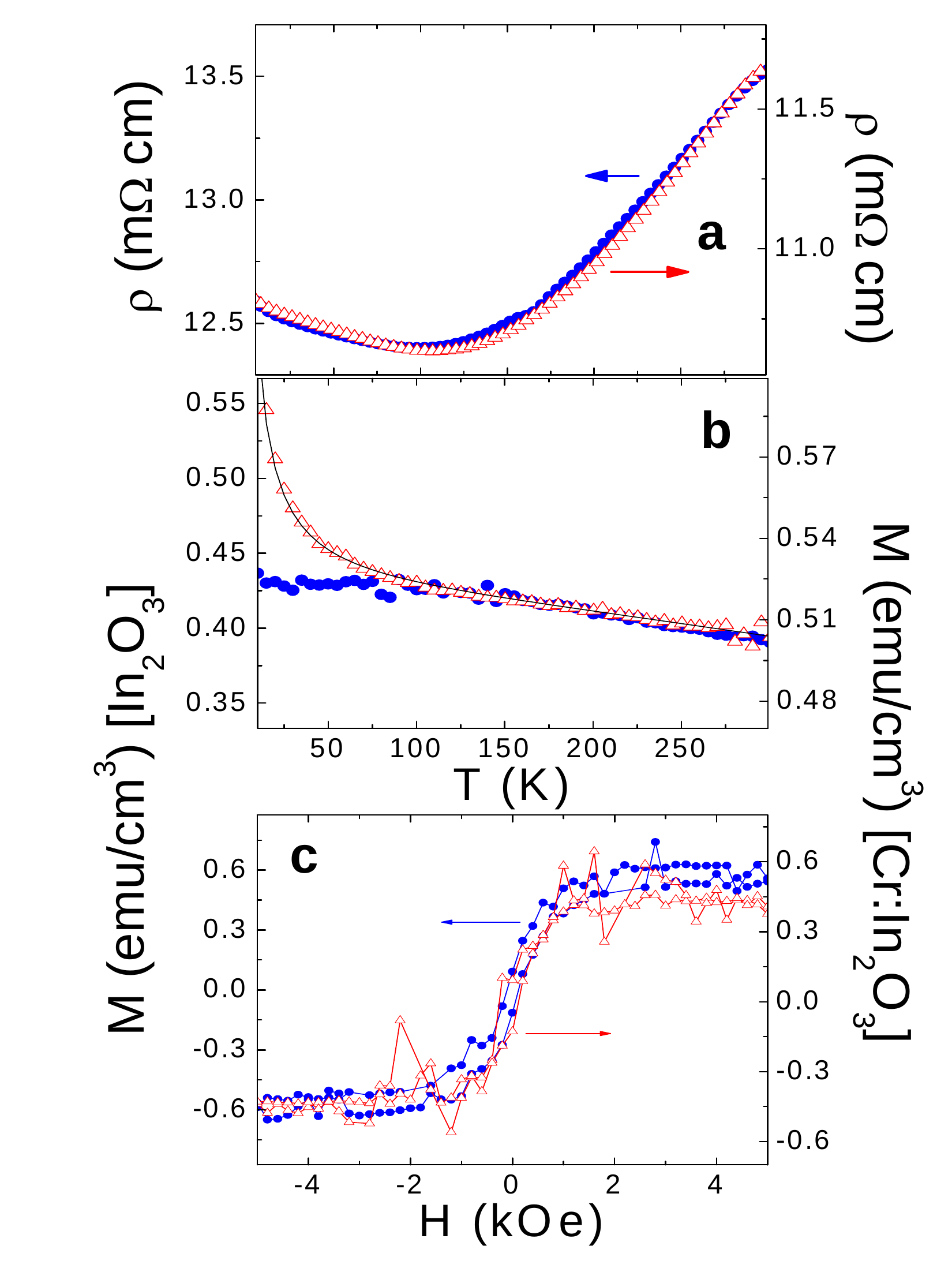}
 \caption{(a)Resistivity versus temperature for the vacuum annealed In$_2$O$_3$\, and 
Cr:In$_2$O$_3$\, samples. (b) Magnetization measured as a
function of temperature for a In$_2$O$_3$\, film and a Cr:In$_2$O$_3$\,
film, measured in a magnetic field of 1 kOe. The solid line is the fit to a
Curie impurity tail, as described in the text. (c) Magnetization versus
magnetic field measured at T=300 K for a In$_2$O$_3$\, film and a Cr:In$_2$O$%
_3$\, film. For all panels, the In$_2$O$_3$\, data are shown with blue
circles and the Cr:In$_2$O$_3$\, data with red triangles.}
\label{fig:mag}
\end{figure}

We plot the temperature dependent electrical resistivity for both the In$_{2}
$O$_{3}$\thinspace\ and Cr:In$_{2}$O$_{3}$\thinspace\ samples in Fig. 
\ref{fig:mag}a. These data were obtained on 1.1 $\mu $m thick samples after
vacuum annealing.  Room
temperature Hall measurements estimate the carrier concentration 
to be 6.1$\times $10$^{19}$cm$^{-3}$ for the In$_{2}$O$_{3}$ \thinspace\
films and  3.5$\times $ 10$^{20}$cm$^{-3}$ for the
Cr:In$_{2}$O$_{3}$\thinspace\ films. Both the In$_{2}$O$_{3}$\thinspace\ and 
Cr:In$_{2}$O$_{3}$\thinspace\ films remain conductive down to low temperatures, and exhibit qualitatively
identical behavior.  This indicates that the electronic properties of both these films are very 
similar.

We measured the in-plane magnetization of these thin film samples using a
high-sensitivity Quantum Design MPMS magnetometer. These magnetic data have
been corrected for a small diamagnetic background from the sapphire
substrate. The magnetization of the as-prepared samples was negligible, but
a sizeable magnetic moment developed on vacuum annealing. We plot the
temperature dependent magnetization for the vacuum annealed In$_{2}$O$_{3}$
\thinspace and Cr:In$_{2}$O$_{3}$\thinspace\ films in Fig. \ref{fig:mag}b.
The magnetizations for both samples exhibit very similar behavior and are
almost temperature independent at higher temperatures, with the 
Cr:In$_{2}$O$_{3}$\thinspace\ sample showing a noticeable increase only at temperatures
below 25 K. We attribute this upturn to a Curie tail arising from
paramagnetic Cr ions in the sample, which are absent in the In$_{2}$O$_{3}$
\thinspace\ film. Fitting this upturn to a Curie susceptibility plus a
spin-wave term as:

\begin{equation}
M(T)=M_{0}+C/T+M_{S}\left( 1-BT^{3/2}\right)   \label{eq_fit}
\end{equation}

\noindent
with M$_{0}$ a constant background, $C$ a Curie term, and $B$ the spin-wave
stiffness, as shown in Fig. \ref{fig:mag}b, we find that this anomaly can be
accounted for by $\approx $75\% of the Cr ions remaining paramagnetic. This
is consistent with our observation that ferromagnetic order can develop in
oxygen deficient In$_{2}$O$_{3}$\thinspace\ in the absence of any magnetic
dopants. The Curie temperature estimated from the fit using Eq.~(\ref{eq_fit}) 
is approximately T$_C$=630 K. We plot the room temperature magnetization curves
for vacuum annealed In$_{2}$O$_{3}$\thinspace\ and Cr:In$_{2}$O$_{3}$\thinspace\ 
films in Fig. \ref{fig:mag}c. Both samples exhibit clear
hysteresis loops, consistent with room temperature ferromagnetic order. Both
the In$_{2}$O$_{3}$\thinspace\ and Cr:In$_{2}$O$_{3}$\thinspace\ films show
a saturation magnetization of approximately 0.5$\pm $0.1 emu/cm$^{3}$.

In order to investigate the coupling between the charge carriers and the
ferromagnetic moment in the In$_{2}$O$_{3}$\thinspace and Cr:In$_{2}$O$_{3}$\thinspace\ films 
we used PCAR spectroscopy to probe the spin polarization.
 PCAR spectroscopy\cite{R3,R4} has recently
emerged as a viable technique to directly measure the transport spin
polarization in magnetic materials\cite{R2,R5} including various magnetic
oxides\cite{R6}, as well as the dilute magnetic semiconductors\cite{R7,R8}. 

\begin{figure} \smallskip \centering 
\includegraphics[width=70mm]{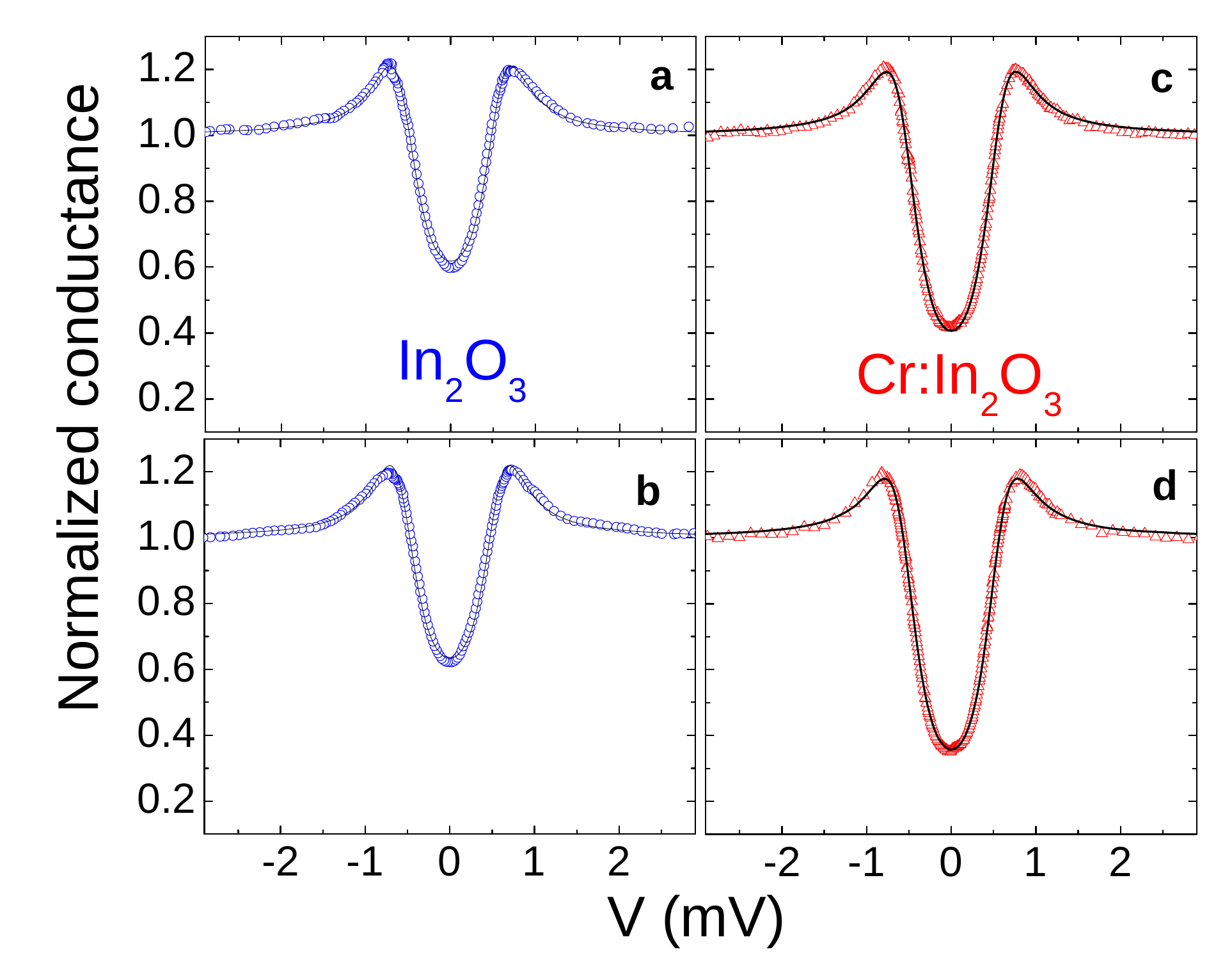}
\caption{(a) and (b) Normalized conductance curves for superconducting Sn
contacts with In$_2$O$_3$\,films (open blue circles) at $T$=1.3 K. (c) and
(d) Normalized conductance curves for superconducting Sn contacts with Cr:In$%
_2$O$_3$\, films (open red triangles) at $T$=1.3 K. Solid lines indicate
numerical fits obtained using the diffusive model{\protect\cite{R10}}. The
BCS gap of bulk Sn, $\Delta$=0.57 meV was used in all the fits. (a) Contact
resistance, R$_c$=87 $\Omega$; fitting parameters: Z=0 and P=42\%. (b) R$_c$%
=75 $\Omega$; fitting parameters: Z=0 and P=40\%. (c) R$_c$=48 $\Omega$,
fitting parameters: Z=0.44 and P=47\%. (d) R$_c$=35 $\Omega$; fitting
parameters: Z=0.47 and P=52\%.}
\label{fig:PCAR}
\end{figure}

All of the measured Sn/In$_{2}$O$_{3}$\thinspace\ and Sn/Cr:In$_{2}$O$_{3}$
\thinspace\ contacts exhibit characteristic conductance curves, with the dip
at zero bias voltage indicating the suppression of Andreev reflection to
spin polarization of the current. Figure \ref{fig:PCAR} shows representative
conductance curves for two different contacts for both samples. The data are
analyzed using the actual BCS gap of bulk Sn, which at the measurement
temperature of T=1.3 K is approximately 0.57 meV. As the typical spreading
resistance of the films - in the range of 20-40 $\Omega$ at 2 K- is comparable to
the point contact resistance, which has an upper limit of $\approx $100 Ohm,
this additional contribution has been included in our analysis \cite{R12}.
We have estimated the minimum interfacial barrier strength value, $Z=\frac{|r-1|}{2\sqrt{r}}$, 
based on the Fermi velocity mismatch between the
superconductor ($v_{Sn}$) and Cr:In$_{2}$O$_{3}$\thinspace\ ($v_{IO}$), $r=%
\frac{v_{Sn}}{v_{IO}}$. Assuming a free electron gas model and taking the
effective mass to be $\approx $0.3 m$_{e}$ \cite{moodera} and the measured
electron density $n\approx 3.5\times 10^{20}$ cm$^{-3}$; the Fermi velocity
of Cr:In$_{2}$O$_{3}$\thinspace\ is calculated $v_{IO}\approx 0.84\times
10^{8}$ cm/s, whereas $v_{Sn}$ is approximately $1.88\times 10^{8}$ cm/s
resulting in the estimated values of $Z$ of $\approx 0.4$, in agreement with
the experimental data for Cr:In$_{2}$O$_{3}$\thinspace\ samples. 

\begin{figure} \smallskip \centering 
\includegraphics[width=70mm]{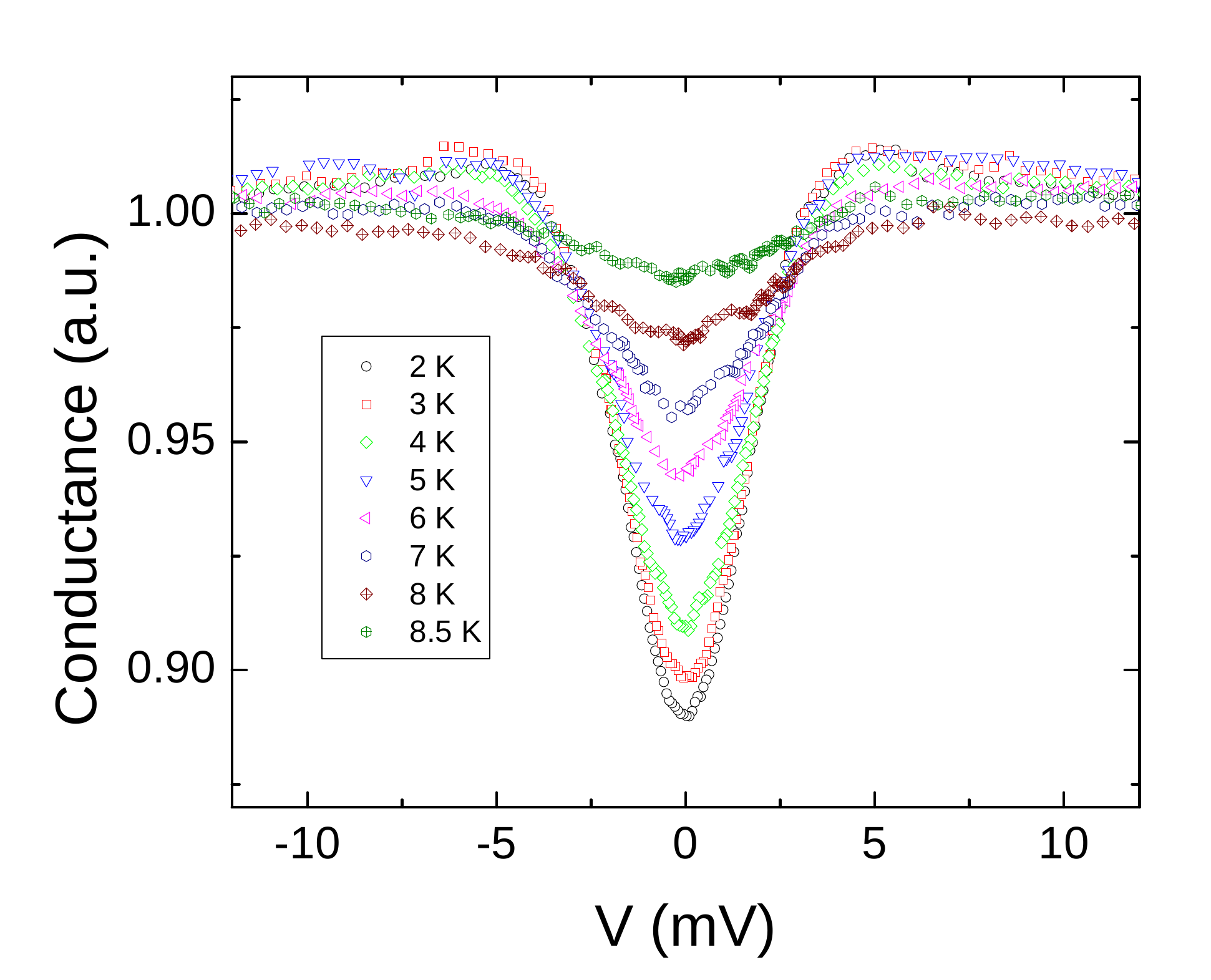}
\caption{
Normalized conductance curve for the \io\, film measured at different temperatures using a Nb
superconducting tip.}
\label{fig:PCAR_T}
\end{figure}

Using the free electron approximation we have estimated the mean free path 
$L\approx 2.5\mathring{A}$ for the resistivity measured at 
$T$=2 K, 10.8 m$\Omega $cm for the Cr:In$_{2}$O$_{3}$\thinspace\ sample. Using the same 
resistivity value and a typical value for the
contact resistance of 50 Ohm, we estimate the contact size $d$ \cite{R11} to
be two orders of magnitude larger than $L$. This implies that $d\gg L$ so
that all of our measurements have been done in the pure diffusive regime.
Accordingly, we have used the diffusive limit of Ref. \cite{R10} to analyze
our data. This analysis, averaged over a number of different point contacts
in several samples, yields a spin polarization of $45\%\pm 5\%$ and $\approx
50\%\pm 5\%$ in the \io\, and \cio\, samples respectively. 
In order to investigate the temperature
dependence of the spin polarization at low temperatures, we have also
performed a series of measurements in In$_{2}$O$_{3}$ at different
temperatures using a Nb tip. The data, shown in Figure \ref{fig:PCAR_T},
 show that the magnitude of the zero bias dip decreases as the superconducting
transition temperature for Nb is approached (Fig. \ref{fig:PCAR_T}). From these
measurements, we calculated 
a spin polarization of $\approx 50\%$ at T=2 K in good agreement with the 
results obtained using the Sn tip.  This value of the spin polarization is approximately
independent of temperature up to T=8 K. As the temperature approaches T$_c$ of Nb ($\approx$ 9K) the 
fluctuations become too large to fit reliably.

The observation of carrier-mediated ferromagnetism coexisting with n-type
conductivity in In$_{2}$O$_{3}$\thinspace\ poses a serious theoretical
challenge. While it has been established that oxygen vacancies are the most
abundant intrinsic (donor) defects in In$_{2}$O$_{3}$ \thinspace and can
account for its room-temperature conductivity \cite{Zunger1}, they alone are
unlikely to produce ferromagnetism.\cite{Zunger2} Recent theoretical
calculations\cite{zunger3} may explain the development of magnetic order
in \cio\, but do not explain the magnetism in undoped \io. It has been proposed that
cation vacancies could be responsible for the observed ferromagnetic
properties of nonmagnetic oxides.\cite{Zunger2,Sawatsky,Sanvito} These
defects have a tendency to form high spin magnetic states that maintain
ferromagnetic interaction at relatively large distances. The spin value of the cation vacancy depends on the
charge state of the defect, with neutral (q=0) and negatively charged (q=-1)
indium vacancies forming S=3/2 and S=1 (triplet) states respectively.
The effect is quite similar to what has been observed in thin 
films of liquid oxygen where two out-of surface oxygen orbitals form triplet states, which
interact ferromagnetically. \cite{oxygenB}  

These assumptions  suggest the following schematic picture. We
propose that vacuum annealed In$_{2}$O$_{3}$ has both oxygen and indium
vacancies. The former act as donors and supply electrons to the conduction
band, while the latter act as acceptors with a localized spin. Free
electrons from oxygen vacancies will mediate an interaction between the
triplet indium vacancies. Spin-ordering on the indium vacancy sites will
further split the conduction band, increasing the free carrier density until
all the donors are ionized. At our measured carrier concentrations in 
In$_{2} $O$_{3}$ on the order of 6$\times $10$^{19}$cm$^{-3}$ the number of
oxygen vacancies would be approximately 3$\times $10$^{20}$cm$^{-3}$ in
good agreement with the theoretical estimates \cite{Zunger1}. The indium
vacancies in In$_{2}$O$_{3}$ would act as compensating defects for the
oxygen vacancy donor states. It is known that In$_{2}$O$_{3}$\thinspace\ is
highly compensated, with the concentration of the free carriers to the donor
defects of about 1:5 \cite{compensation} instead of the expected 2:1 \cite{Zunger1}. 
This compensation suggests that the In vacancy concentration
could also be on the order of a few percent, which, as we will show below,
may be sufficient to explain the onset of ferromagnetic order above room temperature in 
these samples. Specifically, let us assume
that In$_{2}$O$_{3}$ is self-doped with donors (oxygen vacanices) with
density $N_{d}$ and self-compensated by magnetic acceptors (indium
vacancies) of density $N_{a}\lesssim N_{d}$. Due to the latter inequality
all the acceptors are negatively charged and carry a spin $J=1$. The
interaction between electron spins $\vec{s}_{i}$ and localized acceptor
spins $\vec{J}_{j}$ is 
\begin{equation}
H_{ex}=-\Gamma _{ex}\sum_{i,j}\delta (\vec{r}_{i}-\vec{R}_{j})\vec{s}%
_{i}\cdot \vec{J}_{j},  \label{eq:kondo}
\end{equation}%

\noindent
where $\Gamma _{ex}$ is the exchange coupling and
 $\vec{r}_{i}$ $(\vec{R}_{j})$ is the position of the carrier (acceptor). 
 At high temperatures, near
the Curie point, all the donors are ionized and the neutrality
condition for the free electron density $n=N_{d}^{+}-N_{a}^{-}$ can be
simplified as $n\equiv N_{d}\nu =N_{d}-N_{a}$ where we introduce the fraction
of free electrons per donor, $\nu$. The ordering of these localized spins will
split the conduction band and lead to a non-zero spin polarization of the
conduction electrons. 

The mean-field magnetizations (in units of $\mu _{B}$)
of the localized spins $m$ and free electrons $s$ are related by: 

\begin{equation}
m=JB_{J}\left[ \frac{J\Gamma _{ex}ns}{k_{B}T}\right] ,  \label{eq:m_mft}
\end{equation}%
\begin{equation}
s=\frac{1}{2}\tanh \left[ \frac{\Gamma _{ex}N_{a}m}{2k_{B}T}\right] .
\label{eq:s_mft}
\end{equation}%

\noindent
where $B_{J}$ is the Brillouin function and we assume in Eq.~(\ref{eq:s_mft}%
) that the free electrons are non-degenerate. By expanding Eqs.~(\ref{eq:m_mft}) 
and~(\ref{eq:s_mft}) for small $s$ and $m$
we obtain an expression for the transition temperature: 

\begin{equation}
k_{B}T_{C}=\Gamma _{ex}\sqrt{N_{a}nJ(J+1)/{12}}=\Gamma _{ex}N_{d}\sqrt{\nu
(1-\nu )/{6}}  \label{eq:TC}
\end{equation}%

\noindent
If we use the value of $\Gamma_{ex}$=0.15~eV$\cdot $nm$^{3}$ 
reported for GaMnAs\cite{ohno1999a} and
using $N_{d}$=2$\times$10$^{21}$~cm$^{-3}$with $\nu =0.2$,  
we obtain $T_{C}$ $\approx $570~K. This rough estimate is in good agreement with the Curie
temperture obtained from Fig. \ref{fig:mag}b ($\approx $630 K), although further
theoretical and experimental studies will be needed to test this qualitative
model.

In conclusion, we have established that vacuum annealed In$_{2}$O$_{3}$%
\thinspace\ thin films exhibit
ferromagnetic order, and that the charge carriers exhibit a sizeable spin
polarization at T=1.3 K. Our direct measurement of the spin polarization
using PCAR spectroscopy shows a spectrum characteristic of the interface
between a superconductor and ferromagnet, with a spin polarization of $%
\approx $ 50\% for these samples. The close agreement between both the magnetic and
transport measurements for \io\, and \cio\, 
\thinspace\ strongly suggests that the presence of magnetic transition metal dopant ions is not 
necessary to produce carrier mediated ferromagnetism in this system. The observation of a
finite spin polarization points to a strong coupling between the charge
carriers and the ferromagnetic moment. This study confirms one of the
principal assumptions underlying the study of room-temperature
ferromagnetism in dilute magnetic semiconducting oxides, namely that the
charge carriers themselves are spin polarized. 

This work was supported by the National Science Foundation under NSF CAREER
DMR-06044823 and NSF CAREER ECS-0239058, by DARPA through ONR Grant
N00014-02-1-0886, by ONR Grant N00014-06-1-0616, by the Institute for
Materials Research at Wayne State University, and by the Jane and Frank
Warchol Foundation.

\end{document}